\documentclass[5p,sort&compress,times,fleqn]{elsarticle}
\usepackage{graphicx}
\usepackage{hyperref}
\usepackage{color}
\usepackage{latexsym}
\usepackage{amssymb}

\newcommand{\be}{\begin{equation}}
\newcommand{\ee}{\end{equation}}
\newcommand{\ba}{\begin{eqnarray}}
\newcommand{\ea}{\end{eqnarray}}
\def\bs{\begin{subequations}}
\def\es{\end{subequations}}

\usepackage{tikz}
\newcommand*\circled[1]{\tikz[baseline=(char.base)]{
            \node[shape=circle,draw,inner sep=1pt] (char) {#1};}}

\def\de{\delta}
\def\g{\gamma}

\def\k{\kappa}

\def\De{\Delta}

\def\s{\sigma}

\def\cF{\mathcal{F}}

\def\B{\Box}
\newcommand{\Eq}[1]{(\ref{#1})}

\def\cob{\color{blue}}

\newcommand{\books}[4]{{#1}, #2, #3, #4}
\newcommand{\oarX}[1]{\href{http://arxiv.org/abs/#1}{{\ttfamily\cob arXiv:#1}}}
\newcommand{\arX}[1]{\href{http://arxiv.org/abs/#1}{{\ttfamily\cob arXiv:#1}}}
\newcommand{\doin}[6]{\href{http://dx.doi.org/#1}{{\cob #2 #3 {#4} (#6) #5}}}
\newcommand{\doinn}[5]{\href{http://dx.doi.org/#1}{{\cob #2 {#3} (#5) #4}}}
\newcommand{\doij}[5]{\href{http://dx.doi.org/#1}{{\cob #2 #3 (#5) #4}}}

\newcommand{\ndoinn}[5]{\href{#1}{{\cob #2 {#3} (#5) #4}}}

\newcommand{\tia}[1]{{#1},}
\def\rme{e}
\def\rmd{d}

\begin{document}

\begin{frontmatter}

\title{Black-hole stability in non-local gravity}

\author{Gianluca Calcagni}
\ead{g.calcagni@csic.es}
\address{Instituto de Estructura de la Materia, CSIC, Serrano 121, 28006 Madrid, Spain}

\author{Leonardo Modesto}
\ead{lmodesto@sustc.edu.cn}
\address{Department of Physics, Southern University of Science and Technology, Shenzhen 518055, China}

\author{Yun Soo Myung}
\ead{ysmyung@inje.ac.kr}
\address{Institute of Basic Science and Department of Computer Simulation, Inje University, Gimhae 50834, Korea}

\begin{abstract}
We prove that Ricci-flat vacuum exact solutions are stable under linear perturbations in a new class of weakly non-local gravitational theories finite at the quantum level. %We also show that Schwarzschild black holes can form by gravitational collapse.
\end{abstract}

% keywords: Non-local gravity; Black holes

\end{frontmatter}

\date{March 22, 2018}

%%%%%%%%%%%%%%%%%%%%%%%%%%%%%%%
%%%%%%%%%%%%%%%%%%%%%%%%%%%%%%%

\section{Introduction}

A popular motivation to study quantum gravity is that a gravitational interaction governed by the laws of quantum mechanics may avoid spacetime singularities \cite{CQC}. A class of proposals aiming at such result and that raised in the last few years is known under the umbrella name of non-local quantum gravity (see \cite{CQC,MoRa4} for reviews and references therein). Non-local quantum gravity is a perturbative field theory of gravitation whose dynamics is characterized by form factors (operators in kinetic-like terms) with infinitely many derivatives. The quantum theory has the good taste of being unitary and renormalizable or finite, for certain choices of form factors. However, the fate of classical singularities is still under debate. For instance, the classical theory seems able to resolve the big-bang singularity and replace it with a bounce \cite{BMS,CMNi} but singular solutions such as a universe filled with radiation are possible \cite{yaudong}.

The case of black holes is especially interesting because, currently, there are three different views: black holes in non-local gravity may form but are singular \cite{CaMo3,MyPa}, they form and are regular \cite{MMN,FZdP,Fro15,FrZe1,FrZe3,Fro16,Ed18b,KMM1}, or they may not even form as asymptotic, classically stable states \cite{BMM,ZZMB}. The first view is the subject of the present paper (we will comment about the second view in the concluding section). Although there is no complete proof of black-hole stability in non-local theories known to be renormalizable, this proof exists for their non-renormalizable counterparts without the Ricci-tensor--Ricci-tens\-or term \cite{CaMo3}. Here we fill this gap and show analytically that black holes may form and be stable in a wide class of theories in the absence of Weyl--Weyl terms,\footnote{In general, a Weyl--Weyl term $C_{\mu\nu\s\tau}\cF_3(\B)\,C^{\mu\nu\s\tau}$ in the action \cite{BCKM,LiMR} may restrict the theory further. The space of solutions is shrunk \cite{KSKS} with respect to its span in the absence of this extra term \cite{CMN3} and, in particular, Schwarzschild singular solutions are forbidden there \cite{KMM1}.} encompassing most of the proposals whose renormalizability has been studied so far \cite{Kra87,kuzmin,Tom97,Mod11,CaMo2,TBM,MoRa4,KSKMR}. This result should not be taken as a negative trait of non-local quantum gravity, for the simple reason (among others we will discuss later) that the control of divergences in the quantum theory goes much beyond the naive removal of all \emph{classical} singularities.

In section \ref{sec1}, we introduce a new theory of non-local gravity where the Laplace--Beltrami operator $\B$ is replaced by the Lichnerowicz operator $\Delta_{\rm L}$. This change does not modify perturbative renormalizability of the theory with respect to previous proposals but it allows one to address the stability issue (section \ref{sec3}) of Ricci-flat solutions (section \ref{sec2}) exactly. An extension of the theory to the electromagnetic sector is discussed in section \ref{sec4}, while section \ref{sec5} is devoted to conclusions.

Our conventions are the following. The metric tensor $g_{\mu \nu}$ has signature $(- + \dots +)$ and the curvature tensors are defined as $R^{\mu}_{\ \nu \rho \sigma} = - \partial_{\sigma} \Gamma^{\mu}_{\nu \rho} + \dots $, $R_{\mu \nu} = R^{\rho}_{\ \mu  \rho \nu}$ and $R = g^{\mu \nu} R_{\mu \nu}$. Terms quadratic in the Ricci tensor or scalar but not in the Riemann tensor will be denoted as $O({\bf Ric}^2)$.

%%%%%%%%%%%%%%%%%%%%%%%%%%%%%%%
%%%%%%%%%%%%%%%%%%%%%%%%%%%%%%%

\section{A new class of non-local gravity theories}\label{sec1}

A general class of theories compatible with unitarity and super-renormalizability or finiteness has the following structure in $D$ dimensions:
\ba
\hspace{-.8cm} S = \frac{1}{2\kappa^2}\!\!\int\!\! \rmd^D x \sqrt{|g|} \left[R + R \cF_0 (\Delta_{\rm L}) R + R_{\mu\nu} \cF_2(\Delta_{\rm L}) R^{\mu\nu} + V_g\right],\!\!\!\label{action}
\ea
where the `potential' term $V_g$ is at least cubic in the curvature and at least quadratic in the Ricci tensor and $\cF_{0,2}$ are form factors, functions of the Lichnerowicz operator $\Delta_{\rm L}$. When acting on a rank-2 symmetric tensor,
\ba
&& \hspace{-0.83cm}
\Delta_{\rm L} X_{\mu\nu} = 2 R^\s\,_{\mu\nu\tau} \, X^\tau\,_\s + R_{\mu \s} \, X^\s\,_\nu + R_{\s\nu} \, X^\s\,_\mu  - \Box X_{\mu\nu} \nonumber \\
&& = - 2 R_{\mu\s\nu\tau} \, X^{\s\tau} + R_{\mu\s} \, X^\s\,_\nu + R_{\s\nu} \, X^\s\,_\mu  - \Box X_{\mu\nu}\,.
\ea
On the trace $X^\mu_\mu$ or on a scalar $X$, $\Delta_{\rm L}X=-\Box X$.

The Lagrangian \Eq{action} is proposed here for the first time but it has all the same properties of the theories considered in \cite{Kra87,kuzmin,Tom97,Mod11,CaMo2,MoRa4,TBM,KSKMR}, in particular, perturbative unitarity and finiteness. The main difference with respect to previous literature is that $\cF_{0,2}$ are functions of $\Delta_{\rm L}$ instead of the Laplace--Beltrami operator $\B$. On a flat background, these operators coincide. Tree-level unitarity is not affected by the Riemann and Ricci tensors present in the form factors because, when we expand the action to second order in the graviton perturbation, such tensors do not give contributions to the propagator. Indeed, the form factors are inserted between two Ricci tensors or scalars that are already linear in the graviton around Minkowski space. Therefore, the Lichnerowicz operator can only affect vertices in Feynman diagrams, but the power-counting analysis of \cite{kuzmin,Tom97} still holds. One has only to replace the variation of the $\Box$ operator with the variation of $\Delta_{\rm L}$.

%%%%%%%%%%%%%%%%%%%%%%%%%%%%%%%
%%%%%%%%%%%%%%%%%%%%%%%%%%%%%%%

\section{Exact Ricci-flat solution}\label{sec2}

Let us recall the proof that any Ricci-flat spacetime (Sch\-warz\-schild, Kerr, and so on) is an exact solution in a large class of super-renormalizable or finite gravitational theories at least quadratic in the Ricci tensor \cite{yaudong}. This calculation was done with the Laplace--Beltrami operator $\B$ but we adapt it to the $\B\to-\De_{\rm L}$ case straightforwardly.

 The equations of motion (EOM) in a compact notation \cite{Mirzabekian:1995ck} for the action (\ref{action}) read
\ba
\hspace{-0.8cm}
E_{\mu\nu} \hspace{-0.2cm} &:=& \hspace{-0.2cm} \frac{ \delta \left[  \sqrt{|g|} \left(R + R\cF_0(\Delta_{\rm L}) R+R_{\alpha \beta} \cF_2(\Delta_{\rm L}) R^{\alpha \beta} + { V_g}\right) \right]}{\sqrt{|g|} \delta g^{\mu\nu}} \nonumber \\
\hspace{-0.2cm} &=& \hspace{-0.2cm} G_{\mu\nu} - \frac{1}{2} g_{\mu\nu} R \cF_0(\Delta_{\rm L}) R -  \frac{1}{2} g_{\mu\nu} R_{\alpha \beta} \cF_2(\Delta_{\rm L}) R^{\alpha \beta}\nonumber\\
&&\hspace{-0.2cm} +2\frac{\delta R}{\delta g^{\mu \nu}} \cF_0(\Delta_{\rm L})R+ \frac{\delta R_{\alpha \beta}}{\delta g^{\mu \nu}} \cF_2(\Delta_{\rm L})R^{\alpha \beta}  \nonumber \\
&& \hspace{-0.2cm} + \frac{\delta R^{\alpha \beta}}{\delta g^{\mu \nu}  } \cF_2(\Delta_{\rm L})R_{\alpha \beta} +  \frac{\delta \Delta_{\rm L}^r}{\delta g^{\mu\nu} }
 \left[\frac{\cF_0(\Delta_{\rm L}^l)-\cF_0(\Delta_{\rm L}^r)}{\Delta_{\rm L}^r - \Delta_{\rm L}^l} R R \right]\nonumber \\
&& \hspace{-0.2cm} 	+ \frac{\delta \Delta_{\rm L}^r}{\delta g^{\mu\nu}}\left[
  \frac{ \cF_2(\Delta_{\rm L}^l)- \cF_2(\Delta_{\rm L}^r)}{\Delta_{\rm L}^r - \Delta_{\rm L}^l} R_{\alpha \beta} R^{\alpha \beta} \right]+ \frac{\delta {V_g}}{ \delta g^{\mu \nu}}=0\,,\label{EOM}
\ea
where $\Delta_{\rm L}^{l,r}$ act on, respectively, the left and right arguments (on the right of the incremental ratio) inside the brackets.

Replacing the \emph{Ansatz} $R_{\mu\nu} =0$ in the equations of motion (\ref{EOM}), the following chain of implications holds in vacuum:
\ba
R_{\mu\nu} =0  \quad \Longrightarrow \quad E_{\mu\nu} = 0 \quad %\nonumber\\
%&\Longrightarrow&  \mbox{Schwarzschild is an exact solution}\nonumber\\
 \Longleftrightarrow&  \frac{\delta {V_g}}{\delta g^{\mu \nu}}= O({\bf Ric})\,.
\ea
In particular, the Schwarzschild metric, the Kerr metric and all the known Ricci-flat metrics in vacuum Einstein gravity are exact solutions of the non-local theory.

%%%%%%%%%%%%%%%%%%%%%%%%%%%%%%%
%%%%%%%%%%%%%%%%%%%%%%%%%%%%%%%

\section{Stability}\label{sec3}

In this section, we study the stability of Ricci-flat solutions under linear perturbations. We focus on the minimal finite theory of gravity compatible with super-renormalizability. Namely, we select $\cF_0= - \cF_2/2$ (we also redefine $\cF_2 \equiv \gamma$) in \Eq{action} and (\ref{EOM}). Tree-level unitarity requires
\be\label{formfactor}
\gamma (\Delta_{\rm L}) = \frac{\rme^{H(\Delta_{\rm L})} - 1}{- \Delta_{\rm L}} \,,
\ee
where $H$ is an analytic function that can be expanded in a series with infinite convergence radius. This type of `gentle' non-locality is called weak and is discussed elsewhere. We will not make use of the form factor \Eq{formfactor} until later.

At quadratic order in the Ricci tensor, the EOM read
\be
G_{\mu\nu}+2 \frac{\delta R_{\alpha \beta}}{\delta g^{\mu \nu}} \gamma (\Delta_{\rm L}) G^{\alpha \beta} + O({\bf Ric}^2) =0\,. \label{EOM1}
\ee
Notice that the omitted higher-order curvature term is at least quadratic in the Ricci tensor but not in the Riemann tensor. This property is crucial for the proof of stability.

To this purpose, we only need to keep terms at most linear in the Ricci tensor in the EOM. Using the variation
\ba
\frac{\delta R_{\alpha\beta}}{\delta g^{\mu \nu}  }=\frac{1}{2}g_{\alpha(\mu}g_{\nu)\beta}\Box+\frac{1}{2}g_{\mu\nu}\nabla_\alpha
\nabla_\beta-g_{\alpha(\mu|}\nabla_\beta\nabla_{|\nu)} \, , \nonumber
\ea
we can rewrite (\ref{EOM1}) as
\ba
0 &=& G_{\mu\nu}+ 2 \Big[ \frac{1}{2} g_{\alpha (\mu} g_{\nu)\beta}\Box
+ \frac{1}{2} g_{\mu\nu} \nabla_\alpha \nabla_\beta
 \nonumber \\
&& -\frac{1}{2}\left(g_{\alpha \mu} \nabla_\beta \nabla_\nu + g_{\alpha \nu} \nabla_\beta \nabla_\mu \right)
\Big] \gamma(\Delta_{\rm L} ) G^{\alpha \beta}+ O({\bf Ric^2}) \nonumber\\
&=&
G_{\mu\nu}+ \Box \gamma(\Delta_{\rm L}) G_{\mu\nu}
\underbrace{+g_{\mu\nu} \nabla_\alpha \nabla_\beta \, \gamma(\Delta_{\rm L}) G^{\alpha \beta}}_{\circled{1}}
\nonumber \\
&& \underbrace{-2\nabla_\beta \nabla_{(\mu}\,\gamma(\Delta_{\rm L} ) \, G_{\nu)}\,^\beta}_{\circled{2}}+ O({\bf Ric^2})\,.
\label{EOM2}
\ea
In the appendix we prove the following non-trivial identity up to Ricci-square terms:
\be
\nabla^\mu \left[\gamma( \Delta_{\rm L} ) G_{\mu\nu} \right]= O({\bf Ric^2})\,.
\label{FigaIdentity}
\ee
Using this expression, one immediately finds that
\ba
\circled{1} = O({\bf Ric^2})\,. \nonumber
\ea
Also, the commutator of covariant derivatives on a symmetric tensor is
\be
 [\nabla_\beta, \nabla_\mu] \, X^{\alpha \beta} =
R^\alpha\,_{\lambda \beta \mu} \, X^{\lambda \beta} + R_{\lambda\mu}\, X^{\lambda \alpha} \, .
\label{commutatori}
\ee
Plugging \Eq{FigaIdentity} and \Eq{commutatori} with $X^{\alpha \beta} = \gamma(\Delta_{\rm L}) \, G^{\alpha \beta}$ (which is linear in the Ricci tensor) into $\circled{2}$, up to operators quadratic in the Ricci tensor one has
\ba
\hspace{-.8cm}\circled{2} \!\!\!&=&\!\!\! -\left( g_{\alpha \mu} \nabla_\beta \nabla_\nu + g_{\alpha \nu} \nabla_\beta \nabla_\mu \right)
 \gamma(\Delta_{\rm L} ) G^{\alpha \beta} \nonumber  \\
\hspace{-.8cm}\!\!\!&=&\!\!\! -\left( g_{\alpha \mu} \nabla_\nu \nabla_\beta + g_{\alpha \mu}  [\nabla_\beta, \nabla_\nu]
+ g_{\alpha \nu} \nabla_\mu \nabla_\beta + g_{\beta \mu}  [\nabla_\beta, \nabla_\mu]
\right) \nonumber\\
&&\!\!\!\times \gamma(\Delta_{\rm L} ) G^{\alpha \beta}  \nonumber \\
\hspace{-.8cm}\!\!\!&=&\!\!\! -g_{\alpha \mu} R^\alpha\,_{\lambda \beta \nu} \, X^{\lambda \beta}
- g_{\alpha \nu} R^\alpha\,_{\lambda \beta \mu} \, X^{\lambda \beta} + O({\bf Ric^2})  \nonumber \\
\hspace{-.8cm}\!\!\!&=&\!\!\! -R_{\mu \lambda \beta \nu} \, X^{\lambda \beta}
- R_{\nu \lambda \beta \mu} \, X^{\lambda \beta} + O({\bf Ric^2})
\nonumber \\
\hspace{-.8cm}\!\!\!&=&\!\!\! 2 R_{\mu \beta \nu \lambda} X^{\beta \lambda} + O({\bf Ric^2})\nonumber
\,.
\ea
Therefore, the EOM ({\ref{EOM2}) simplifies to
\ba
0&=& G_{\mu\nu}+ \Box \gamma(\Delta_{\rm L}) G_{\mu\nu}
+ 2 R_{\mu \alpha \nu \beta} \gamma(\Delta_{\rm L}) G^{\alpha \beta}  + O({\bf Ric^2}) \nonumber\\
&=&
[1-\Delta_{\rm L} \gamma(\Delta_{\rm L})] G_{\mu\nu}+O({\bf Ric^2})\,,
\label{EOM4}
\ea
which agrees with the full equations of motion \cite{CMN2}. Replacing now the form factor \Eq{formfactor} in (\ref{EOM4}), we get
\be
\rme^{H(\Delta_{\rm L})} G_{\mu\nu} + O({\bf Ric^2})=0\,. \label{LastEOM}
\ee
Provided $\exp H$ is entire (all renormalizable non-local quantum gravities satisfy this property\footnote{Contrary to the typical non-analytic form factors $\g\propto\ln(-\B)$ appearing in the effective quantum action coming from a local field theory \cite{BarV1,BarV2,GuZ,GoSh1,GoSh2,Sha08,BDFM}.}), we obtain the equations for linear perturbations around any Ricci-flat background:
\ba
\rme^{H(\Delta_{\rm L})} \delta G_{\mu\nu}=0 \quad \Longrightarrow \quad \delta G_{\mu\nu}=0 \,,\label{Pert}
\ea
where the last implication is clear in a basis in which the Lichnerowicz operator is diagonal. Note that, at the linearized level, the number of degrees of freedom of nonlocal gravity is two; the other six degrees of freedom \cite{CMN2,CMN3} are fully non-perturbative and, consistently, they do not show up in \Eq{Pert}.

The conclusion is that the problem of linear perturbations in nonlocal gravity can be reduced to exactly the same problem in Einstein gravity. Expanding the metric $g_{\mu\nu}=\bar{g}_{\mu\nu}+h_{\mu\nu}$ around a Ricci-flat background $\bar{g}_{\mu\nu}$ with a small perturbation $h_{\mu\nu}$ ($|h_{\mu\nu}|\ll 1$), equation \Eq{Pert} is equivalent to solve $\delta R_{\mu\nu}(h)=0$. Any Ricci-flat spacetime stable in Einstein gravity is stable in non-local gravity, too. In particular, the Schwarzschild metric is stable because it is stable in general relativity, since the perturbation $h_{\mu\nu}$ remains small throughout its evolution \cite{ReWi,Vis70,Zer70,Wal79,KW,KI1,KI2,Gui06,DHR}. Explicitly, from $\delta R_{\mu\nu}(h)= (\bar{\nabla}^{\rho}\bar{\nabla}_{\mu}h_{\nu\rho}+
\bar{\nabla}^{\rho}\bar{\nabla}_{\nu}h_{\mu\rho}-\bar{\nabla}^2h_{\mu\nu}-\bar{\nabla}_{\mu}\bar{\nabla}_{\nu}h)/2=0$, the perturbation $h_{\mu\nu}$ is classified depending on the transformation properties under parity, namely, odd and even. Using the Regge--Wheeler--Zerilli gauge, one obtains two distinct types of perturbations: odd with 2 degrees of freedom and even with 4 degrees of freedom. We have to mention that even though one starts with $2+4=6$ degrees of freedom in the Regge--Wheleer--Zerilli gauge, the physically propagating degrees of freedom are still two of odd and even type for a massless spin-2 mode (a gauge-invariant treatment is also available \cite{KI1,KI2}). This result mainly concerns the stability on and outside the horizon. We do not consider the case inside the horizon because the situation is quite different near the singularity inside the black hole.

Note that this result is much simpler than in higher-order local gravity \cite{Whi85} because the derivative operator in \Eq{Pert} is an entire function, while in the higher-order local case one must solve a second-order equation for $\delta R_{\mu\nu}(h)$.

%%%%%%%%%%%%%%%%%%%%%%%%%%%%%%%
%%%%%%%%%%%%%%%%%%%%%%%%%%%%%%%

\section{Non-local Einstein--Maxwell gravity}\label{sec4}

In this section, we propose an extension of the non-local theory that could admit charged black holes as solutions. Based on previous results on completely decoupled non-local gauge (in particular, electromagnetic) sectors \cite{Modesto:2015foa,MoRa2}, a natural way to couple non-local gravity with a non-local Maxwell field is via a Lagrangian of the following form:
\ba
&& \hspace{-1cm}
\mathcal{L} =  \frac{1}{2\kappa^2} \left[ R + \left(G_{\mu\nu}-\k^2\tau^{A}_{\mu\nu}
\right)
\cF_{\rm g}^{\mu\nu, \, \rho \sigma} (\Delta_{\rm L}) \left(G_{\rho \sigma}-\k^2\tau^{A}_{\rho \sigma}\right) \right] \nonumber \\
&& \hspace{0.3cm}
- \frac{1}{4} F_{\mu\nu}F^{\mu\nu} +  \nabla_\mu F^{\mu\nu} \, \cF_A(\Delta_{\rm L})  \, \nabla_\rho F^{\rho}_{\ \nu},
\label{GT}
\ea
where the analytic functions of the Lichnerowicz operator $\cF_{g}$, $\cF_A$ and the rank-2 tensor $\tau^{A}_{\mu\nu}$ are defined as
\ba
&& \hspace{-1cm}
\cF_{g}^{\mu\nu, \,\rho \sigma}(\Delta_{\rm L}) := \left(g^{\mu \rho} g^{\nu\sigma} - \frac{1}{2} g^{\mu \nu} g^{\rho \sigma} \right) \left( \frac{\rme^{H_{\rm g} (\Delta_{\rm L})} -1}{-\Delta_{\rm L}} \right)  \, , \nonumber \\
&&\hspace{-1cm}
\cF_{A}(\Delta_{\rm L}) :=  \frac{\rme^{H_{A} (\Delta_{\rm L}) } -1}{-2\Delta_{\rm L}}\,, \\
&&  \hspace{-1cm}
\tau^{A}_{\mu\nu} := F_{\mu\sigma} F^{\sigma}_\nu - \frac{1}{4} F_{\mu\nu}F^{\mu\nu} \, , \quad
F_{\mu\nu} := \partial_\mu A_{\nu} - \partial_\nu A_{\mu} \nonumber\,,
\ea
where $A_\mu$ is Abelian. We propose the theory (\ref{GT}) as a viable coupling of quantum gravity with the electromagnetic force. Its main properties can be guessed from the results of \cite{Modesto:2015foa,MoRa2} and it may be argued to be ghost-free in the gravitational and electromagnetic sectors separately. The polynomial asymptotic ultraviolet scaling of the entire functions $H_g$ and $H_A$ must be the same in order to have the same fall-off of the propagator in the ultraviolet regime and a renormalizable theory.  A formal proof of these statements will be given elsewhere.

When we take the variation of the Lagrangian (\ref{GT}) with respect to the metric, we find an EOM very similar to (\ref{EOM}) but in the presence of the electromagnetic field. In particular, the operator $\propto \cF_{\rm g}^{\mu\nu, \, \rho \sigma}$ is quadratic in the local Einstein EOM, while the operator $\propto \cF_{\rm A}$ is quadratic in the EOM for $A_{\mu}$. As a consequence, if the local EOM are satisfied, so are the non-local EOM:
\ba
&& G_{\rho \sigma} = \k^2 \, \tau^{A}_{\rho \sigma}   \quad \Longrightarrow \quad E_{\mu\nu} = \k^2 T_{\mu\nu} \, , \nonumber \\
&& \nabla_\mu F^{\mu\nu} = 0   \qquad \Longrightarrow \quad E_A^{\mu} = 0 \,,\label{IGA}
\ea
where $E_A^\mu:=\de S/\de A_\mu$ is not calculated explicitly here. The local EOM are on the left-hand side of the conditionals (in particular, the electromagnetic field is described by two-derivative Maxwell theory), while the exact non-local EOM for the metric and the electromagnetic field are on the right-hand side. Note that when the EOM $G_{\rho \sigma} = \k^2 \, \tau^{A}_{\rho \sigma}$ are satisfied, we have $T_{\mu\nu} =\tau^{A}_{\rho \sigma}$, which means that $\tau^{A}_{\rho \sigma}$ becomes the physical energy-momentum tensor  on shell. Therefore, all the solutions of the Einstein EOM in the presence of the electromagnetic field are also solutions of non-local gravity. The latter may have more solutions but the point here is that all black-hole solutions of general relativity, including those incorporating an electromagnetic charge, are exact solutions of the theory (\ref{GT}). Schwarz\-schild, Kerr and Reissner--Nordstr\"om metrics are all admissible.

If one coupled non-local gravity and electromagnetism to normal matter, then the\, Reissner--Nordstr\"om\, solution\, would prob\-a\-bly not be an exact solution anymore. Indeed, the form factors in the theory can delocalize (smear) the source and the energy-momentum tensor will be modified everywhere in spacetime. Therefore, it remains to be seen whether the solution we found is physical.

%Let us now come back to Vaidya gravitational collapse. The Maxwell-gravity action \Eq{GT} falls short of describing a null dust because a null dust made of neutral massless particles may be regarded as an incoherent superposition of electromagnetic plane waves, while the field theory \Eq{GT} describes individual plane waves of photons. However, an educated guess is that switching to incoherent radiation would lead to an effective action similar to \Eq{GT}, where $\tau^{A}_{\mu\nu}$ is replaced by $\tau^{\rm ND}_{\mu\nu}$, the second line of \Eq{GT} is set to zero and ultraviolet divergences are under control when the energy-momentum tensor can be derived by a fundamental theory similar to the one we just proposed for the photon. Doing this, the Vaidya metric (\ref{Vaidya_metric}) with the particular energy-moment\-um tensor (\ref{Tmunu}) is an exact solution of the non-local theory (\ref{GT}) and we conclude that the Schwarzschild metric can be created by the gravitational collapse of a thin shell of radiation.

%%%%%%%%%%%%%%%%%%%%%%%%%%%%%%%
%%%%%%%%%%%%%%%%%%%%%%%%%%%%%%%

\section{Conclusions}\label{sec5}

In this paper, we completed the stability analysis of Ricci-flat spacetimes in a large class of non-local gravity. In previous works, it was proved that Minkowski and any maximally symmetric spacetime are stable under linear perturbations. In this paper, we extended the analysis of \cite{CaMo3} of Ricci-flat spacetimes for non-renormalizable models to renormalizable theories. We showed that the stability analysis for a particular class of non-local gravitational theories reduces exactly to the one in Einstein gravity, i.e., the linearized equation $\de R_{\mu\nu}(h)=0$. Therefore, the analysis for Einstein gravity \cite{ReWi,Vis70,Zer70,Wal79,KW,KI1,KI2,Gui06,DHR} holds automatically also in non-local gravity and one can conclude that the Schwarzschild black hole is stable under linear perturbations, which means that the latter are either bounded or suppressed. Our method is very general because, as just said, we brought back the stability problem in non-local gravity to the same problem in Einstein's theory. For instance, one can immediately borrow the results of Kerr black holes for the non-local case. The Kerr solution is stable outside the outer horizon \cite{Whi89,FKSY}\footnote{Note that mode stability \cite{Whi89} excludes a particular type of growing solutions, hence it is different from linear stability, where all solutions are bounded or decay.} but unstable inside the inner horizon \cite{DGRV}. Extending the gravitational action to the electromagnetic sector, we argued that the Reissner--Nordstr\"om metric is also a solution, although we have not checked its stability.

Naively, one might regard the presence of singular solutions in a non-local setting as a support to the claim \cite{yaudong} that non-local quantum gravity may not provide a resolution of the singularity problem, even when the quantum theory is finite. However, one cannot reach the conclusion that singular physical black holes exist in the full theory. Indeed, a finite quantum theory can hide a conformal invariance both at the classical and the quantum level \cite{ll16,Modesto:2018def}. It may be the latter, rather than non-locality itself, that solves the spacetime singularity problem \cite{cuta8,Bambi:2016wdn,Bambi:2016yne,Bambi:2017ott,Bambi:2017yoz,Chakrabarty:2017ysw}.

Moreover, numerous solutions of the truncated theory \cite{MMN} or of the linearized theory \cite{FZdP,Fro15,FrZe1,Fro16,FrZe3} point out that, in the presence of matter, we could have non-singular and non-Ricci-flat black holes.\footnote{See also \cite{FrZe2,BFZ} for other regular sources. Regular black holes were also found in effective quantum-gravity models with non-local operators of the form $\ln\B$ \cite{FrVi1,FrVi2}, to be contrasted with the fundamental non-locality we are dealing with in this paper, which has different ultraviolet properties.} In particular, these results suggest that a Ricci-flat black hole may be not a physical metric resulting from the collapse of ordinary matter. However, in non-local gravity no Birkhoff theorem has been formulated yet and it is not obvious how to interpret our spherically symmetric Ricci-flat solutions in vacuum, which are \emph{exact} in the fully \emph{non-linear} theory, in relation with spherically symmetric non-Ricci-flat approximate solutions of the \emph{linearized} equations of motion with matter. The next steps to take will be, on one hand, to check whether the stable solutions found with our analysis are physical and, on the other hand, to apply the same analysis also to regular black holes in the spacetime region where deviations from the Ricci-flat metric are small.

%%%%%%%%%%%%%%%%%%%%%%%%%%%%%%%
%%%%%%%%%%%%%%%%%%%%%%%%%%%%%%%

\section*{Acknowledgments}

\noindent G.C.\ is under a Ram\'on y Cajal contract and thanks SUSTech for the kind hospitality during the writing of this work. G.C.\ and L.M.\ are supported by the MINECO I+D grants FIS2014-54800-C2-2-P and FIS2017-86497-C2-2-P.

%%%%%%%%%%%%%%%%%%%%%%%%%%%%%%%
%%%%%%%%%%%%%%%%%%%%%%%%%%%%%%%

\appendix

\section{Proof of (\ref{FigaIdentity})}

In this appendix, we prove the identity (\ref{FigaIdentity}). Consider first the covariant derivative of one Lichnerowicz operator acting on the Einstein tensor,
\ba
\hspace{-.8cm} \nabla^\mu \left( \Delta_{\rm L} G_{\mu\nu} \right)\!\!\!  &=&\!\!\! - \nabla^\mu \left( \Box G_{\mu\nu} + 2 R_{\mu \rho \nu \sigma} G^{\rho \sigma} \right)   +O({\bf Ric^2})   \nonumber \\
\hspace{-.8cm} \!\!\! &=& \!\!\!- \Big[ \underbrace{\nabla^\mu \nabla_\alpha \nabla^\alpha G_{\mu\nu} }_{\circled{I}}
+  \underbrace{2 \nabla^\mu \left(  R_{\mu \rho \nu \sigma} G^{\rho \sigma}  \right)}_{\circled{II}} \Big]\nonumber\\
&&+O({\bf Ric^2})\,.\label{step1}
\ea
We know from the Bianchi identities that
\ba
\hspace{-1.4cm}&&\nabla^\mu G_{\mu\nu} = 0\,,\\
\hspace{-1.4cm}&&-\nabla^\alpha R_{\nu \sigma \beta \alpha} = \nabla_\nu R_{\sigma \beta} - \nabla_\sigma R_{\nu \beta} \,,\label{contractedBianchi} \\
\hspace{-1.4cm}&&[\nabla_\rho,\nabla_{\mu_1}]X^{\mu_1\mu_2\mu_3} = R^{\mu_1}_{\ \lambda \rho \mu_1} X^{\lambda \mu_2 \mu_3}
+ R^{\mu_2}_{\ \lambda \rho \mu_1} X^{\mu_1 \lambda \mu_3}\nonumber\\
\hspace{-1.4cm}&&\hspace{2.6cm}+ R^{\mu_3}_{\ \lambda \rho \mu_1} X^{\mu_1 \mu_2 \lambda}\,.\label{perb2}
\ea
Therefore,
\ba
&& \hspace{-0.98cm}
\circled{I} =  \nabla_\alpha \nabla^\mu  \nabla^\alpha G_{\mu\nu} + [ \nabla^\mu , \nabla_\alpha ] \nabla^\alpha G_{\mu\nu}  \nonumber \\
&& \hspace{-0.5cm} = \nabla_\alpha   \nabla^\alpha  \nabla^\mu G_{\mu\nu}
+  \nabla_\alpha  [ \nabla^\mu,  \nabla^\alpha ]  G_{\mu\nu}
+  [ \nabla^\mu , \nabla_\alpha ] \nabla^\alpha G_{\mu\nu} \nonumber \\
&&  \hspace{-0.5cm}
= \underbrace{\nabla_\alpha  [ \nabla^\mu,  \nabla^\alpha ]  G_{\mu\nu}}_{\circled{a}}
+  \underbrace{[ \nabla^\mu , \nabla_\alpha ] \nabla^\alpha G_{\mu\nu}}_{\circled{b}} \, .
\label{step2}
\ea
We can express $\circled{a}$ in terms of the Riemann tensor,
\ba
&&
\hspace{-0.8cm}
\circled{a} = \nabla^\alpha  [ \nabla^\mu,  \nabla_\alpha ]  G_{\mu\nu}
=  \nabla^\alpha \Big( \underbrace{R^\mu\,_{\lambda \mu \alpha}}_{\bf Ric} G^\lambda\,_\nu
- R^\lambda\,_{\nu \mu \alpha} G^\mu\,_\lambda \Big) \nonumber \\
&& \hspace{-0.3cm} =
-  (\nabla^\alpha R^\lambda\,_{\nu \mu \alpha})  G^\mu\,_\lambda
- R^\lambda\,_{\nu \mu \alpha}   \nabla^\alpha G^\mu\,_\lambda + O({\bf Ric^2})\nonumber \\
&& \hspace{-0.3cm} =- R_{\alpha \mu \nu \lambda}   \nabla^\alpha  G^{\mu \lambda} + O({\bf Ric^2}) \,,
\label{step4}
\ea
where in the last line we used (\ref{contractedBianchi}). Also, from (\ref{perb2}) we get
\ba
[ \nabla_{\rho}, \nabla_{\mu_1}] \nabla^{\mu_1} G^{\rho \mu_3}
= R^{\mu_3}\,_{\lambda \rho \mu_1} \nabla^{\mu_1} G^{\rho \lambda } + O({\bf Ric^2})\nonumber
\ea
and
\ba
&&\hspace{-0.5cm}
 \circled{b}
= g_{\delta \nu}  [\nabla_\mu , \nabla_\alpha] \nabla^\alpha G^{\mu \delta}
= g_{\delta \nu}  R^\delta\,_{\lambda \mu \alpha}\nabla^\alpha G^{\lambda \mu} +O({\bf Ric^2})  \nonumber \\
&&
= - R_{\alpha \mu \nu \lambda}  \nabla^\alpha G^{\lambda \mu}   +O({\bf Ric^2})\,.
\label{bbb}
\ea
Replacing (\ref{step4}) and (\ref{bbb}) in (\ref{step2}), we obtain
\ba\nonumber
\circled{I} = -2 R_{\alpha \mu \nu \lambda}  \nabla^\alpha G^{\mu \lambda} + O({\bf Ric^2})\,.
\ea

The operator $\circled{II}$ can be simplified using (\ref{contractedBianchi}):
\ba
\hspace{-0.8cm} \circled{II} &=& 2 \nabla^\mu \left( R_{\mu \rho \nu \sigma} G^{\rho \sigma}  \right) = 2 \big(\underbrace{\nabla^\mu  R_{\mu \rho \nu \sigma}}_{{\bf Ric}} \big) G^{\rho \sigma}  + 2 R_{\mu \rho \nu \sigma} \nabla^\mu G^{\rho \sigma} \nonumber\\
\hspace{-0.8cm} &=& 2 R_{\mu \rho \nu \sigma} \nabla^\mu G^{\rho \sigma} + O({\bf Ric^2}) \,,\nonumber
\ea
so that $\circled{I} + \circled{II}=O({\bf Ric^2})$ and
\be
\nabla^\mu \left( \Delta_{\rm L} G_{\mu\nu} \right) = O({\bf Ric^2})\,.
\ee

Introducing the symmetric tensor $Z_{\mu\nu} := \Delta_{\rm L} G_{\mu\nu}$ satisfying the property $\nabla^\mu Z_{\mu\nu}  = O({\bf Ric^2})$, in exactly the same way as before we can show that
\ba
\nabla^\mu \left(\Delta_{\rm L} Z_{\mu\nu} \right) =\nabla^\mu \left(\Delta_{\rm L} \Delta_{\rm L} G_{\mu\nu} \right)  = O({\bf Ric^2}) \,.\nonumber
\ea
Similarly,
\be
\nabla^\mu \left( \Delta_{\rm L} \Delta_{\rm L} \cdots \Delta_{\rm L} G_{\mu\nu} \right)  = O({\bf Ric^2})\,,
\ee
and finally, for an analytic function of the Lichnerowicz operator, we obtain \Eq{FigaIdentity}.

%%%%%%%%%%%%%%%%%%%%%%%%%%%%%%%
%%%%%%%%%%%%%%%%%%%%%%%%%%%%%%%

\end{document}